\documentclass[aps,pra]{revtex4}
\usepackage{epsf,epsfig,graphics,graphicx}
\begin{document}

\title{Quasi-exactly solvable quasinormal modes}
%\classification{03.65.-w, 03.65.Nk, 03.65.Fd, 04.70.-s}
%\keywords{Quasi-exact solvability, quasinormal modes}

\author{\underline{Choon-Lin Ho} and Hing-Tong Cho}
%{address={Department of Physics, Tamkang University,
% Tamsui 25137, Taiwan, R.O.C.} }

\affiliation{Department of Physics, Tamkang University, Tamsui
251, Taiwan, Republic of China}

\begin{abstract}
We consider quasinormal modes with complex energies from the point
of view of the theory of quasi-exactly solvable (QES) models. We
demonstrate that it is possible to find new potentials which admit
exactly solvable or QES quasinormal modes by suitable
complexification of parameters defining the QES potentials.
Particularly, we obtain one QES and four exactly solvable
potentials out of the five one-dimensional QES systems based on
the $sl(2)$ algebra.
\end{abstract}

\maketitle

{\bf Introduction.--}~~Quasinormal modes (QNM) arise as
perturbations of stellar or black hole spacetimes \cite{QNM}. They
are solutions of the perturbation equations that are outgoing to
spatial infinity and the event horizon.  Generally, these
conditions lead to a set of discrete complex eigenfrequencies,
with the real part representing the actual frequency of
oscillation and the imaginary part representing the damping.  QNM
carry information of black holes and neutron stars, and thus are
of importance to gravitational-wave astronomy.  In fact, these
oscillations, produced mainly during the formation phase of the
compact stellar objects, can be strong enough to be detected by
several large gravitational wave detectors under construction.
Recently, QNM of particles with different spins in black hole
spacetimes have also received much attention.

Owing to the intrinsic complexity in solving the perturbation
equations in general relativity with the appropriate boundary
conditions, one has to resort to various approximation methods,
eg., the WKB method, the phase-integral method etc., in obtaining
QNM solutions.  It is therefore helpful that one can get some
insights from exact solutions in simple models, such as the
inverted harmonic oscillator and the P\"oschl-Teller potential.
Unfortunately, the number of exactly solvable models is rather
limited.

Recently, in non-relativistic quantum mechanics a new class of
potentials which are intermediate to exactly solvable ones and
non-solvable ones has been found. These are called quasi-exactly
solvable (QES) problems for which it is possible to determine
analytically a part of the spectrum but not the whole spectrum
\cite{TU,Tur,Ush,GKO}.  The discovery of this class of spectral
problems has greatly enlarged the number of physical systems which
we can study analytically.  In the last few year, QES theory has
also been extended to the Pauli and Dirac equations.

In \cite{ChoHo} we have considered solutions of QNM based on the
Lie-algebraic approach of QES theory.  There we demonstrate that,
by suitable complexification of some parameters of the generators
of the $sl(2)$ algebra while keeping the Hamiltonian Hermitian, we
can indeed obtain potentials admitting exact or quasi-exact QNMs.
Such consideration has not been attempted before in studies of QES
theory.  Our work represents a direct opposite of the work in
\cite{BB}, where QES {\it real energies} were obtained from a {\it
non-Hermitian} {\cal PT}-symmetric  Hamiltonian.

{\bf QES Theory.--}~~Let us briefly review the essence of the
Lie-algebraic approach \cite{TU,Tur,GKO} to QES models. Consider a
Schr\"odinger equation $H\psi=E\psi$ with Hamiltonian $H=-d_x^2 +
V(x)$ ($d_x\equiv d/dx$) and wave function $\psi (x)$. Here $x$
belongs either to the interval $(-\infty,\infty)$ or $[0,\infty)$.
Now suppose we make an ``imaginary gauge transformation" on the
function $\psi$: $\psi (x)= \chi(x) e^{-g(x)}$, where $g(x)$ is
called the gauge function. For physical systems which we are
interested in, the phase factor $\exp(-g(x))$ is responsible for
the asymptotic behaviors of the wave function so as to ensure
normalizability. The function $\chi(x)$ satisfies a Schr\"odinger
equation with a gauge transformed Hamiltonian $H_g=e^g H e^{-g}$.
Suppose $H_g$ can be written as a quadratic combination of the
generators $J^a$ of some Lie algebra with a finite dimensional
representation. Within this finite dimensional Hilbert space the
Hamiltonian $H_g$ can be diagonalized, and therefore a finite
number of eigenstates are solvable. Then the system described by
$H$ is QES.  For one-dimensional QES systems the most general Lie
algebra is $sl(2)$, and $H_g$ can be expressed as
\begin{eqnarray}
H_g=\sum C_{ab}J^a J^b + \sum C_a J^a + {\rm real\ constant}~,
\label{H-g}
\end{eqnarray}
 where $C_{ab},~C_a$ are taken to be \emph{real constants} in  \cite{Tur,GKO}.
The generators $J^a$ of the $sl(2)$ Lie algebra  take the
differential forms: $ J^+ = z^2 d_z - nz~,~ J^0=z
d_z-n/2~,~J^-=d_z$ ($n=0,1,2,\ldots$). The variables $x$ and $z$
are related by some function to be described later. $n$ is the
degree of the eigenfunctions $\chi$, which are polynomials in a
$(n+1)$-dimensional Hilbert space with the basis $\langle
1,z,z^2,\ldots,z^n\rangle$.

Substituting the differential forms of $J^a$ into Eq.~(\ref{H-g}),
one sees that every QES operator $H_g$ can be written in the
canonical form : $H_g=-P_4(z) d_z^2 + P_3(z)d_z + P_2(z)$, where
$P_k (z)$ are $k$-th degree polynomial in $z$ with real
coefficients related to the constants $C_{ab}$ and $C_a$. The
relation between $H_g$ and the standard Schr\"odinger operator $H$
fixes the required form of the gauge function $g$ and the
transformation between the variable $x$ and $z$. Particularly,
$x=\int^z dy/\sqrt{P_4 (y)}$.  Analysis on the inequivalent forms
of real quartic polynomials $P_4$ thus give a classification of
all $sl(2)$-based QES Hamiltonians \cite{Tur,GKO}.  If one imposes
the requirement of non-periodic potentials, then there are only
five inequivalent classes, which are called case 1 to 5 in
\cite{GKO}.

Our main observation is this.  If some of the coefficients in $P_k
(z)$ are allowed to be complex while keeping $V(x)$ real, then all
the five cases classified in  \cite{GKO} can indeed support
QES/exact quasinormal modes.  We shall illustrate this using one
of the cases below.

{\bf QES QNM.--}~~We consider Case 3 in \cite{GKO}, which
corresponds to Class I potential in Turbiner's scheme \cite{Tur}.
There are two subclasses in this case, namely, Case (3a) and (3b).
We shall present the analysis of QNM potential for case (3a). The
other case turns out to give the same potential with a suitable
choice of the parameters. The potential in case (3a) has the form
(in this paper we adopt the notation of \cite{GKO}):
\begin{eqnarray}
V(x)=A e^{2\sqrt{\nu} x}+ B e^{\sqrt{\nu} x}+ C e^{-\sqrt{\nu} x}
+ D e^{-2\sqrt{\nu} x}~, \label{V-3a}
\end{eqnarray}
where $x\in (-\infty,\infty)$ and $\nu$ is a positive scale
factor. Note that $V(x)$ is defined up to a real constant, which
we omit for simplicity, as it merely shifts the real part of the
energy. This remark also applies to the other cases.   $V(x)$ in
Eq.~(\ref{V-3a}) reduces to the exactly solvable Morse potentials
when $A=B=0$, or $C=D=0$. This potential is QES when the
coefficients are related by
\begin{eqnarray}
 A&=& \frac{\hat{b}^2}{4\nu}~,~~~
 B=\frac{\hat{c}+(n+1)\nu}{2\nu}~\hat{b},~~~\nonumber\\
 C&=&\frac{\hat{c}-(n+1)\nu}{2\nu}~\hat{d},~~~
 D=\frac{\hat{d}^2}{4\nu},~\\
 &&~n=0,1,2\ldots\nonumber
 \end{eqnarray}
Here $\hat{b},~\hat{c},~\hat{d}$ are arbitrary real constants. For
each integer $n\geq 0$, there are $n+1$ exactly solvable
eigenfunctions in the $(n+1)$-dimensional QES subspace:
\begin{equation}
\psi_n(x)=\exp{\left[\frac{\hat{b}}{2\nu} e^{\sqrt{\nu} x} +
\frac{\hat{c}-n\nu}{2\sqrt{\nu}} x - \frac{\hat{d}}{2\nu}
e^{-\sqrt{\nu} x}\right]}\chi_n (e^{\sqrt{\nu} x})~. \label{psi-n}
\end{equation}
Here $\chi_n (z)$ is a polynomial of degree $n$  in
$z=\exp(\sqrt{\nu}x)$.  To guarantee normalizability of the
eigenfunctions, the real constants
$\hat{b},~\hat{c},~\hat{d},~\nu$ and $n$ must satisfy certain
relations \cite{GKO}.

We want to see if we can get QNM solutions if we allow some
parameters to be complex, while still keeping the potential $V(x)$
real. This latter requirement restricts the possible values of the
parameters, and hence the forms of QES potential admitting
quasinormal modes. For the case at hand, we find that one possible
choice of values of $\hat{b},~\hat{c}$ and $\hat{d}$ is:
\begin{equation}
\hat{b}=ib ,~\hat{c}=-(n+1)\nu,~\hat{d}=d,~~b,d:{\rm real\
constants}~.
\end{equation}
The potential Eq.~(\ref{V-3a}) becomes
\begin{eqnarray}
V_n(x)=-\frac{b^2}{4\nu} e^{2\sqrt{\nu} x}- \left(n+1\right) d
e^{-\sqrt{\nu} x} + \frac{d^2}{4\nu} e^{-2\sqrt{\nu} x}~,
\label{V-3a1}
\end{eqnarray}
and the wave function Eq.~(\ref{psi-n}) becomes
\begin{eqnarray}
\psi_n(x)&=&\exp{\left[\frac{ib}{2\nu} e^{\sqrt{\nu} x} -
\left(n+\frac{1}{2}\right)\sqrt{\nu} x - \frac{d}{2\nu}
e^{-\sqrt{\nu} x}\right]}\nonumber\\ &&~~~~~\times \chi_n
(e^{\sqrt{\nu} x})~. \label{psi-n1}
\end{eqnarray}
$V(x)$ approaches $\mp\infty$ as $x\to\pm\infty$ respectively: it
is unbounded from below on the right. For small positive $d$ and
sufficiently large $n$, $V(x)$ can have a local minimum and a
local maximum. In this case, the well gets shallower as $d$
increases at fixed value of $n$, or as $n$ decreases at fixed $d$.
We emphasize here that for different value of $n$, each $V(x)$
represents a different QES potential admitting $n+1$ QES
solutions. Since $V(x)\to \infty$ as $x\to -\infty$, the wave
function must vanish in this limit. This means $d>0$ from
Eq.~(\ref{psi-n1}). For the outgoing boundary condition, we must
take $b>0$.  Before we go on, we note here that there is another
possible choice of the parameters, namely,
\begin{equation}
\hat{b}=-b ,~\hat{c}=(n+1)\nu,~\hat{d}=id,~~b,d:{\rm real\
constants}~.
\end{equation}
However, this choice leads to a potential related to
Eq.~(\ref{V-3a1}) by the reflection $x\mapsto -x$.  Hence, we will
only discuss the potential in Eq.~(\ref{V-3a1}) here.

To see that the wave functions $\psi_n(x)$ do represent
quasinormal modes, we determine the corresponding energy $E_n$.
This is easily done by solving the eigenvalue problem of the
polynomial part $\chi_n (z)$ of the wave function.  From the
Schr\"odinger equation we find that for $n=0$, the energy is $E_0=
-\nu/4- ibd/2\nu$.  This clearly shows that the only QES solution
when $n=0$ is a QNM with an energy having a negative imaginary
part (recall that $b,d,\nu>0$). For $n=1$, we have two QES
solutions. Their energies are $E_1=-5\nu/4- ibd/2\nu \pm
\sqrt{\nu^2 -ibd}$. Again we have two QNM modes. One can proceed
accordingly to obtain $n+1$ QNM modes with higher values of $n$.
However, for large $n$, computation becomes tedious, and one has
to resort to numerical means.

One can do the same for the other four cases listed in \cite{GKO}.
It is found that these cases admits exact QNM solutions.

{\bf Summary.--}~~To summarize, we have demonstrated that it is
possible to extend the usual QES theory to accommodate QNM
solutions, by complexifying certain parameters defining the QES
potentials. We found that  the five $sl(2)$-based QES systems
listed in \cite{GKO} can be so extended.   While  one of these
cases admits QES QNM, the other four cases give exact QNM
solutions. It is hoped that our work would motivate the search of
many more exact/quasi-exact systems of QNM in QES theories based
on higher Lie algebras, and in higher dimensions.

\vskip 1.5 truecm

This work was supported in part by the National Science Council of
the Republic of China under the Grants NSC 94-2112-M-032-009
(H.T.C.) and NSC 94-2112-M-032-007 (C.L.H.).

\end{document}